\documentclass{ametsocv6.1}

\usepackage[utf8]{inputenc}
\usepackage[utf8]{inputenc}
\usepackage{amsmath}
\usepackage{caption}
\usepackage[T1]{fontenc}
\usepackage{amssymb,amsmath}
\usepackage{graphicx}
\usepackage{gensymb}
\usepackage{url}
\usepackage{enumitem}
\usepackage{longtable}
\usepackage{float}

\newcommand{\co}{CO$_2${}}

\nolinenumbers
\title{\nolinenumbers Exploring subtropical stratocumulus multiple equilibria using a mixed-layer model}

\authors{\nolinenumbers Andrea M. Salazar,\aff{a}\correspondingauthor{\nolinenumbers Andrea M. Salazar, andreasalazar@g.harvard.edu} 
Eli Tziperman\aff{a,b} 
}

\affiliation{\nolinenumbers \aff{a}{Harvard University, \textit{Department of Earth and Planetary Sciences}, Cambridge, MA, USA}
\aff{b}{Harvard University, \textit{School of Engineering and Applied Sciences}, Cambridge, MA, USA}
}
\nolinenumbers

\abstract{Stratocumulus clouds cover about a fifth of Earth’s surface, and due to their albedo and low-latitude location, they have a strong effect on Earth’s radiation budget. Previous studies using Large Eddy Simulations have shown that multiple equilibria (both cloud-covered and cloud-free states) exist as a function of fixed-SST, with relevance to equatorward advected air masses. Multiple equilibria have also been found as a function of atmospheric \co, with a subtropical SST nearly 10 K higher in the cloud-free state, and with suggested relevance to warm-climate dynamics. \\
In this study, we use a mixed-layer model with an added surface energy balance and the ability to simulate both the cloud-covered and cloud-free states to study both types of multiple equilibria and the corresponding hysteresis. The model's simplicity and computational efficiency allow us to explore the mechanisms critical to the stratocumulus cloud instability and hysteresis as well as isolate key processes that allow for multiple equilibria via mechanism denial experiments not possible with a full-complexity model. For the hysteresis in fixed-SST, we find that decoupling can occur due to either enhanced entrainment warming or a reduction in cloud-top longwave cooling. The critical SST at which decoupling occurs is highly sensitive to precipitation and entrainment parameterizations. In the \co{} hysteresis, decoupling occurs in the simple model used even without the inclusion of an active SST and SST-cloud cover feedbacks, and the width of the hysteresis displays the same sensitivities as the fixed-SST case. Overall, the simple model analysis and results motivate further studies using higher complexity models.}

\begin{document}
\maketitle

\section{Introduction}

Stratocumulus clouds are low-altitude clouds that form beneath sharp temperature inversions, where convection between the surface and the cloud layer is driven by cloud-top longwave cooling \citep{turton1987study,bretherton1997moisture,wood2012stratocumulus}. On an annual average, stratocumulus clouds cover about 20\% of Earth's surface and are particularly concentrated over subtropical eastern ocean margins \citep{warren1988global}. Their location, combined with their prevalence and high albedo, means that they have a considerable effect on Earth's radiation budget, providing a net local forcing of up to $-100$ W/m$^2$ in subtropical regions \citep{klein1993seasonal}. 

Cloud-top longwave cooling causes turbulent convective motions that homogenize the boundary layer and couple surface moisture fluxes to the cloud layer. These fluxes sustain the cloud layer against strong entrainment warming and drying through mixing with the inversion layer above \citep{lilly1968models, nicholls1984dynamics, bretherton1997moisture}. Turbulence is also enhanced by latent heat fluxes from the surface and latent heat release within the cloud layer \citep{bretherton1997moisture}. Convection between the surface and cloud layer is interrupted when cloud-top longwave cooling weakens \citep{lilly1968models} or when entrainment warming becomes too strong \citep{bretherton1997moisture}. Under these conditions, the cloud layer is effectively cut off (decoupled) from its surface moisture supply and the stratocumulus clouds give way to much less reflective scattered cumulus clouds. The decrease in albedo associated with this cloud transition can cause considerable surface warming \citep{schneider2019possible}.

\cite{bellon2016stratocumulus} found a hysteresis in stratocumulus cloud cover as a function of fixed-SST using a Large Eddy Simulation (LES), where multiple equilibria (that is, both cloudy and non-cloudy states) are possible between SST values of 288 K and 293 K. The presence of multiple equilibria was attributed to the cloud radiative effect, in which a very dry cloud would have a low emissivity and weak longwave cooling while a moist cloud would have a high emissivity and strong longwave cooling. The demise of the stratocumulus equilibrium with increased SST was attributed to an enhancement of entrainment drying from the inversion layer above the cloud. The enhanced entrainment is driven at high prescribed SST by turbulence produced by increased latent heat fluxes from the surface. This enhanced entrainment drying leads to a lower liquid water path in the cloud layer and therefore weaker longwave cooling at the cloud-top. This fixed-SST scenario ignores the cloud cover-SST feedback.

More recent work has investigated how stratocumulus cloud cover in the presence of an interactive SST can change due to \co-driven climate warming, and experience bistability and hysteresis. \citet{schneider2019possible} analyzed a hysteresis of stratocumulus clouds as a function of atmospheric \co{}. For \co{} values between 300 ppm and 1200 ppm, both cloud-covered (coupled) and cloud-free (decoupled) states were possible. They attributed the breakup of the stratocumulus equilibrium at high \co{} to the increased opacity of the free troposphere at higher \co{}, which increases the downwelling longwave radiation and therefore weakens the net longwave cooling at the cloud-top, ceasing convection. As the \co{} is then decreased, it needs to be lowered to \co{} values smaller than the original critical \co{} value in order to regain the coupled equilibrium. The abrupt breakup of the cloud layer due to rising \co{} levels may be relevant to future climate change as well as to past periods of hothouse climates such as the Eocene (56--48 Myr). This adds to previous work on cloud feedbacks in warm climates, including a possible role for polar stratospheric clouds \citep{Sloan-Pollard-1998:polar, Kirk-Davidoff-Schrag-Anderson-2002:feedback}, a convective Arctic cloud feedback that can lead to multiple equilibria and keep the arctic ocean ice-free \citep{Abbot-Tziperman-2008:sea,abbot2009controls}, and low-clouds over land due to the advection of moist air from the ocean, which can prevent the surface from reaching sub-freezing temperatures during winter time \citep{Cronin-Tziperman-2015:low}.

Stratocumulus topped boundary layers (STBLs) have been studied with multiple models, ranging from simple mixed-layer models (MLM) \citep{lilly1968models, turton1987study, bretherton1997moisture, pelly2001mixed} to Large Eddy Simulations \citep[LES,][]{uchida2010sensitivity, bretherton1997moisture, bellon2016stratocumulus, schneider2019possible}. Simple models have not been used to investigate the hysteresis behavior found in LES as a function of SST or \co{} \citep{schneider2019possible, bellon2016stratocumulus}, because the assumption of a well-mixed boundary layer breaks down when the layer decouples from the surface. However, the diurnal cycle of stratocumulus cloud was studied by \citet{turton1987study}, who used a ``stacked'' mixed-layer approach to model decoupled solutions.

Our objective is to explore the processes behind the multiple equilibria and hysteresis behavior of stratocumulus cloud layers as a function of both \co{} and SST. For this purpose, we adapt a mixed-layer model to include a surface energy balance and the ability to model decoupled solutions. With this model, we reproduce the instability that causes a break-up of stratocumulus clouds with increasing SST following \citet{bellon2016stratocumulus} and in \co{} based on \citet{schneider2019possible}. We then explore (i) what nonlinearities specifically allow for multiple equilibria, and (ii) what factors influence the critical SST or \co{} at which the cloud layer onset or breakup occurs. Using a simple model allows us to both isolate key processes and efficiently explore a wide parameter space.

In Section 2 we describe the modified MLM that includes a surface energy balance and allows decoupled solutions. This model is then used to explore the dynamics of multiple equilibria and hysteresis in \co{} and fixed-SST in Section 3. We conclude in Section 4.

\section{Model Description}
\label{sec:model-description}

\begin{figure}[!ht]
    \centering
    \includegraphics[width = \linewidth]{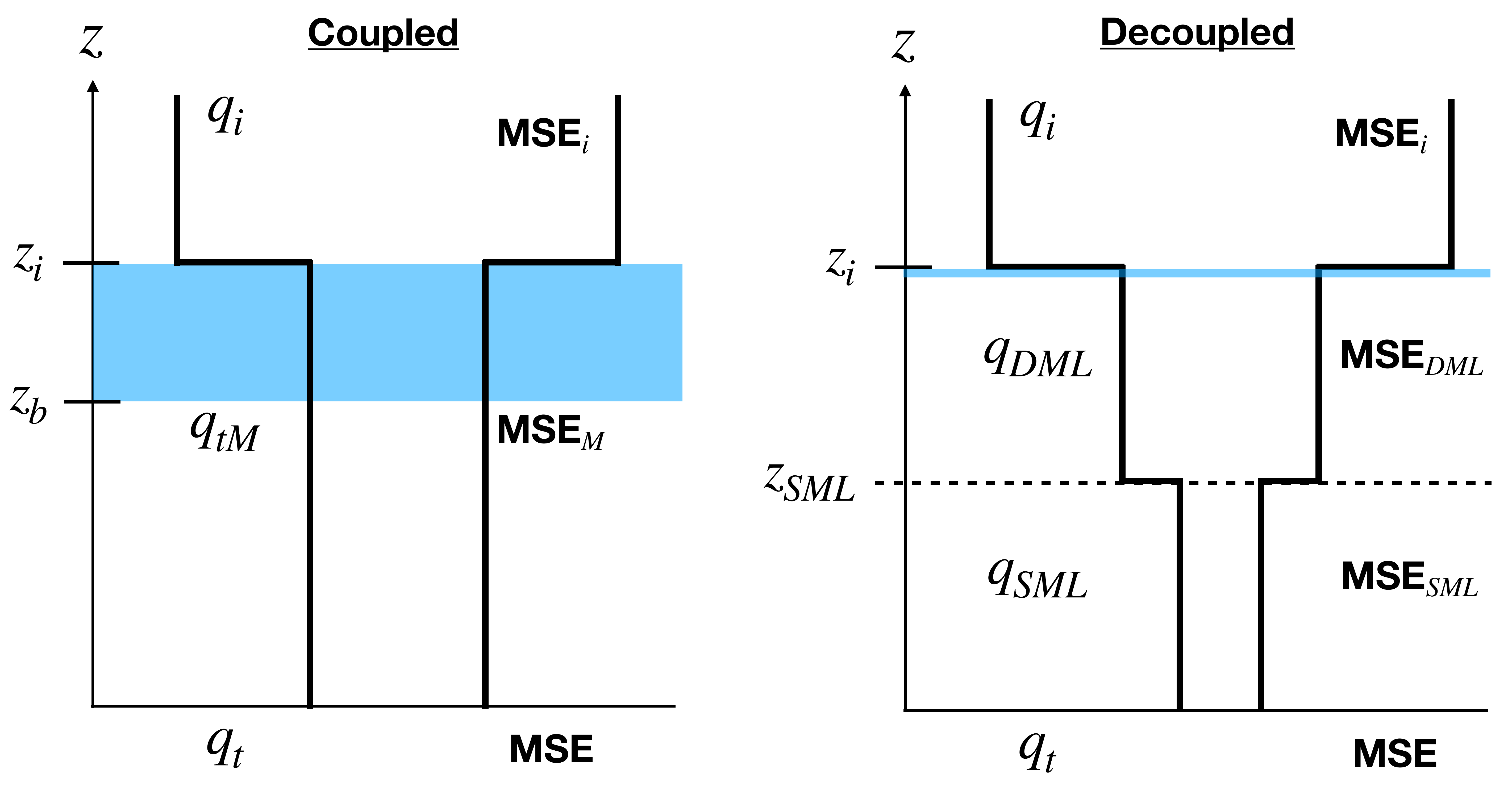}
    \caption{Vertical profiles of total (vapor plus liquid) moisture, $q_t$, and moist static energy, MSE, in both the coupled and decoupled states. The blue patch represents the cloud layer within the mixed-layer which is infinitely thin when the model is decoupled.}
    \label{fig:model_schem}
\end{figure}

In this section, we describe the mixed-layer model including the interactive ocean mixed-layer component. We assume that moist static energy, $\text{MSE}= c_p T + L q_v + g z$, and total (vapor plus liquid) moisture content, $q_t = q_v + q_l$, are well mixed throughout the layer in the coupled state and are therefore constant in height \citep{lilly1968models}. In Section 2\ref{sec:coupled-model}, we describe the coupled mixed-layer model equations that follow \citet{bretherton1997moisture}, with a surface energy balance added. In Section 2\ref{sec:decoupled-model}, we follow \citet{turton1987study} and model decoupled solutions by stacking a ``surface mixed layer'' beneath the ``decoupled mixed layer''. The difference in vertical structure of conserved variables is summarized in Figure \ref{fig:model_schem}.

\begin{figure}[!ht]
    \centering
    \includegraphics[width = \linewidth]{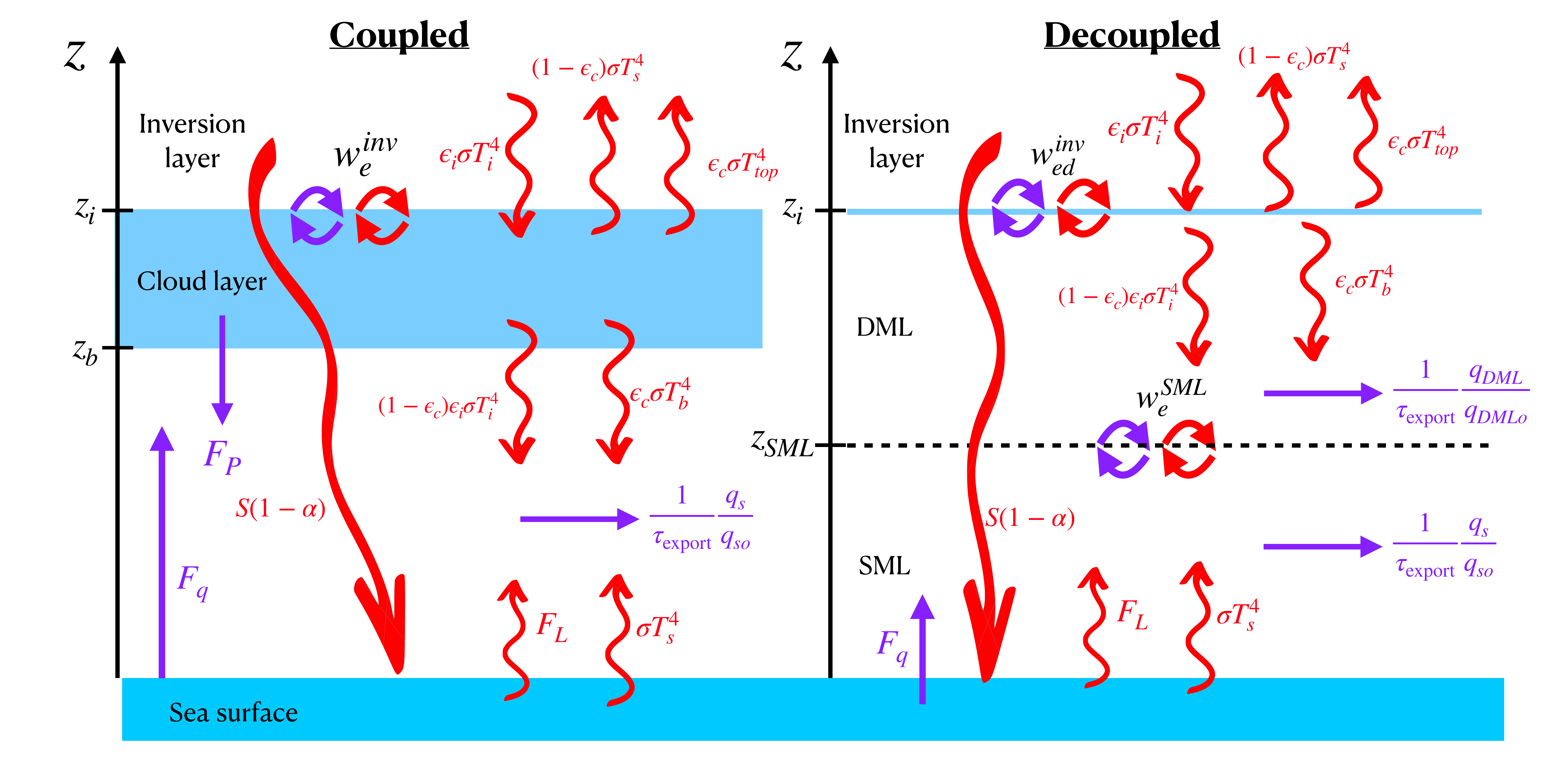}
    \caption{A schematic of the fluxes in the mixed-layer model where red arrows correspond to terms in the energy budgets for the surface and mixed-layer(s). Purple arrows correspond to terms in the moisture budget. When the model is decoupled, the cloud layer is infinitely thin and there is no precipitation.}
    \label{fig:fluxes_schem}
\end{figure}

\subsection{Modelling the coupled state} \label{sec:coupled-model}

As in \citet{bretherton1997moisture}, the prognostic equations for moist static energy in the mixed layer, MSE$_M$, and total moisture in the mixed layer, $q_{tM}$, are
\begin{align}
    \frac{d \text{MSE}_M}{dt} &= -\frac{\partial E}{\partial z}= -\frac{\partial}{\partial z} \left(\left\langle w'\text{MSE}'\right\rangle + F_R/\rho_a\right)
    \label{eq:base_MSE}\\
    \frac{d q_{tM}}{dt} &= -\frac{\partial W}{\partial z}= -\frac{\partial }{\partial z}\left(\left\langle w'q_{tM}'\right\rangle - F_P\right) \label{eq:base_qtM}.
\end{align}
where $E$ and $W$ are the total energy and moisture vertical fluxes. Angle brackets indicate horizontally averaged fluxes and $F_R$ (W/m$^2$) and $F_P$ (kg/kg times m/s) are the net upward radiative flux and the precipitation flux, respectively. The conservation equations for the full mixed layer are found by integrating Equations \eqref{eq:base_MSE} and \eqref{eq:base_qtM} from the surface to the top of the layer, $z_i$.
\begin{table}
\centering
\begin{tabular}[c]{ c| c| l }
\hline
\hline
Variable & Value & Description \\
\hline
$\rho_a$ & 1.225 kg/m$^3$ & Air density \\
$C_T$ & $0.001(1+0.07V)$ & Transport parameter \\ 
 V & 7.1 m/s & 10-m wind \\ 
$T_{io}$ & 290 K & Initial inversion temperature \\ 
$\Gamma_i$ & 6.5 K/km & Lapse rate in inversion layer \\
RH$_i$ & 0.4 & Relative humidity of inversion layer \\
C & 0.5 & Entrainment efficiency \\
$w^{inv}_{ed}$ & 0.0056 m/s & Decoupled cloud-top entrainment, \co-hysteresis \\
$w^{inv}_{ed}$ & 0.002 m/s & Decoupled cloud-top entrainment, SST-hysteresis \\
$\tau_\mathrm{precip}$ & 2 hrs & Precipitation  \\
$\Gamma_p$ & $0.86 \times 10^{-5}$ kg/kg 1/km & Evaporation rate beneath cloud base \\
$\tau_\mathrm{export}$ & 3.96 years & Moisture export  \\
$\epsilon_{c0}$ & 0.45 & Minimum cloud layer emissivity \\ 
D & 1.66 & Diffusivity factor \\
$\kappa$ & 90.36 m$^2$/kg & Absorption coefficient of water droplets \\
$A$ & 0.05 & Climate sensitivity to CO$_2$ \\
$B$ & 0.1 & Climate sensitivity to H$_2$O \\
$\epsilon_{i0}$ & 0.5 & Minimum inversion layer emissivity \\
Div & $6\times 10^{-6}$ 1/s & Large scale subsidence\\
$p_o$ & 1000 hPa & Surface pressure \\
S & 471 W/m$^2$ & Diurnally averaged insolation \\
$\mathrm{OHU}$ & 10 W/m$^2$ & Ocean heat uptake \\
$\alpha_w$ & 0.06 & Albedo of water \\
$\alpha_c$ & 0.6 & Maximum albedo of cloud \\
$\mu$ & 0.93 & Thermodynamic factor (Equation \eqref{eq:DC}) \\
$w^{SML}_e$  & 0.013 m/s & Entrainment in SML, \co-hysteresis \\
$w^{SML}_e$ & 0.003 m/s & Entrainment in SML, SST-hysteresis \\
$q^{DML}_{t0}$  & 8 g/kg & Initial DML total moisture, \co-hysteresis \\
$q^{DML}_{t0}$ & 3.5 g/kg & Initial DML total moisture, SST-hysteresis \\
$z_{SML}$ & 200 m & Top of SML \\
$z_{id}$ & 1000 m & Decoupled inversion height \\
 \hline
\end{tabular}
\caption{Table of constants in order of appearance in the text.}
\end{table}

In order for the MSE and total moisture in the mixed layer to remain vertically uniform, the energy and moisture flux convergences appearing in the above equation must be independent of height and therefore the corresponding fluxes must be linear functions of height,
\begin{align}
    E(z) &= \left(1-\frac{z}{z_i}\right)E(0) + \frac{z}{z_i}E(z_i) \label{eq:E_height}\\
    W(z) &= \left(1-\frac{z}{z_i}\right)W(0) + \frac{z}{z_i}W(z_i).
\end{align}
Surface fluxes, defined at $z=0$, are calculated using bulk aerodynamic formulas,
\begin{align}
    E(0) &= C_TV(\text{MSE}_s - \text{MSE}_M) + F_R(0)/\rho_a \\
    W(0) &= C_TV(q_s-q_{tM}) - F_P(0) \label{eq:W_ground},
\end{align}
where MSE$_s$ and $q_s$ are the saturation surface moist static energy and specific humidity. At the surface, $F_R(0)$ is the net upwards radiation at the surface and $F_P(0)$ is the precipitation that reaches the surface. At the inversion above the mixed layer, entrainment mixing lead to energy and moisture fluxes of the form,
\begin{align}
    E(z_i) &= -w^{inv}_e (\text{MSE}_i - \text{MSE}_M) + F_R(z_i)/\rho_a \\
    W(z_i) &= -w^{inv}_e(q_i - q_{tM}) \label{eq:W_flux_top},
\end{align}
where MSE$_i$ and $q_i$ are the moist static energy and specific humidity in the inversion layer. We assume the warm air in the inversion layer originates from updrafts in the tropics and will warm with increasing \co{} levels \citep{schneider2019possible} assuming a climate sensitivity of 3 \degree{}C. We therefore prescribe the temperature of the inversion layer as a function of \co{} and inversion height,
\begin{equation}
    T_i = T_{io} + 3\log_2\left(\frac{\text{CO}_2}{280}\right) - \Gamma_i(z_i - z_{io}),
\end{equation}
where $T_{io}$ and $z_{io}$ are the temperature and altitude of the inversion layer at 280 ppm. The lapse rate in the inversion layer, $\Gamma_i$, is assumed to follow a moist adiabat. We assume a constant relative humidity of the inversion layer, RH$_i$, so that the specific humidity in the inversion layer can be calculated from the Clausius-Clapeyron relationship.

Following \citet{pelly2001mixed}, we define the entrainment rate, $w^{inv}_{e}$, of warm, dry air from above the inversion into the cloud layer as
\begin{equation}
    \rho_a c_p w^{inv}_{e} =\begin{cases} C\frac{\Delta F_R - \mu L \Delta F_p}{\Delta \theta^{i-c}_v} & \text{Coupled} \\
    \rho_a c_p w^{inv}_{ed} & \text{Decoupled},
    \label{eq:we}\end{cases}
\end{equation}
where $C$ is a prescribed constant entrainment efficiency, $\Delta F_R$ is the net longwave cooling by the cloud layer, $\Delta F_P$ is the net precipitation flux across the mixed-layer, and $\Delta \theta^{i-c}_v$ is the jump in virtual potential temperature across the inversion. Entrainment is driven by convective turbulence in the mixed-layer and is inhibited by a stable stratification across the inversion layer. Therefore, sources and sinks of turbulence (net longwave cooling and evaporative cooling of drizzle beneath the cloud-base) appear in the numerator of Equation \eqref{eq:we}  and the inversion strength is in the denominator. This parameterization is very similar to the energy balance entrainment rate described in \citet{bretherton1997moisture} and the ``minimum entrainment'' closure by \citet{lilly1968models}. When the cloud layer is decoupled from the surface, we set the entrainment to a background value, $w^{inv}_{ed}$. 

The precipitation flux, $F_P$ is assumed constant throughout the cloud and decreases linearly with height beneath the cloud base at a rate of $\Gamma_P$, as in \citet{turton1987study}. By defining a constant timescale of precipitation, we write the precipitation flux as a function of height, $z$, and cloud liquid water path, LWP,
\begin{equation}
    F_p(z) = \begin{cases} \frac{\text{LWP}}{\rho_a \tau_\mathrm{precip}} & z_i > z \ge z_b \\
    \frac{\text{LWP}}{\rho_a \tau_\mathrm{precip}} - \Gamma_p(z_b-z) & z_b < z,
    \end{cases}
  \label{eq:precip_flux}
\end{equation}
where $z_b$ is the cloud-base. 
Combining Equations \eqref{eq:base_MSE}--\eqref{eq:W_flux_top}, we write the time evolution of MSE$_{M}$ and $q_{tM}$ as,
\begin{align}
    z_i\frac{d\text{MSE}_M}{dt} &= C_TV(\text{MSE}_s - \text{MSE}_M) + w^{inv}_e(\text{MSE}_i -\text{MSE}_M) - \Delta F_R/\rho_a \label{eq:MSE_balance_coupled} \\
    z_i\frac{dq_{tM}}{dt} &= C_TV(q_s - q_{tM}) - w^{inv}_e(q_{tM} -q_i) - \Delta F_P - z_i\frac{q_s}{q_{so}}\frac{1}{\tau_\mathrm{export}}. \label{eq:moisture_balance_coupled}
\end{align}
The last term contains an added moisture export out of the subtropics and into other regions of the form used in \citet{schneider2019possible} that scales with the saturation specific humidity at the surface, $q_s$ and has a constant timescale of export, $\tau_\mathrm{export}$ 

The net longwave cooling across the layer, $\Delta F_R$, is the difference in the net upwards longwave flux between the cloud-top and the surface. Since there is no precipitation flux entering from above at the cloud-top, the net precipitation flux across the layer is simply the precipitation flux at the surface, $F_P(0)$. We use a simple radiation parameterization, illustrated in Figure \ref{fig:fluxes_schem}, rather than the two-stream approximation used by \citet{bretherton1997moisture}. At the cloud-top, the net upward longwave radiation is,
\begin{equation}
    F_R(z_i) = \epsilon_c\sigma T^4_{top} + (1-\epsilon_c)\sigma T_s^4 - \epsilon_i \sigma T_i^4 \label{eq:net_flux_zi},
\end{equation}
where $T_{top}$ is the temperature of the cloud top, found by making vertical profiles of temperature using the MSE$_M$ and $q_{tM}$. The cloud base, $z_{b}$, is defined as the altitude where the saturation point is reached. The emissivity of the cloud, $\epsilon_c$, is found as a function of LWP by assuming the cloud acts as a grey body,
\begin{equation}
    \epsilon_c = 1- (1-\epsilon_{co})\exp(-D\kappa \text{LWP}) \label{eq:ec},
\end{equation}
where $D$ is a nondimensional diffusivity factor and $\kappa$ is the absorption coefficient of liquid water \citep{UCAR_cloud_emiss}. In the absence of liquid water (no cloud), the emissivity is set to a minimum value, $\epsilon_{co}$, due to the \co{} and water vapor in the mixed layer.

The emissivity of the inversion layer is written as a function of \co{} and $q_i$,
\begin{equation} \label{eq:inv_emiss}
    \epsilon_i = \epsilon_{io} + A\log_2\left(\frac{\text{CO}_2}{280}\right) + B\log_2\left(\frac{q_i}{q_{io}}\right),
\end{equation}
where $A$ is tuned by running the full MLM model to give a surface warming of 1.5 \degree{}C for a doubling of \co{} when $B = 0$. The value of $B$ is tuned to give a surface warming of 3 \degree{}C for a doubling of \co{} when including the water vapor feedback.

At the surface, the net upwards radiation is,
\begin{equation}
    F_R(0) = \sigma T_s^4 - \epsilon_c\sigma T_{b}^4 - (1-\epsilon_c)\epsilon_i \sigma T_i^4 \label{eq:net_flux_0},
\end{equation}
where $T_{b}$ is the temperature at the base of the cloud. Combining Equations \eqref{eq:net_flux_zi} and \eqref{eq:net_flux_0}, the net longwave cooling across the mixed layer is,
\begin{equation}
    \Delta F_R = \epsilon_c\sigma\left(T_{top}^4 + T_{b}^4\right) - \epsilon_c \sigma T_s^4 - \epsilon_c\epsilon_i\sigma T_i^4 \label{eq:net_flux_tot}.
\end{equation}

Finally, we write the time evolution of the inversion height, $z_i$, and the surface temperature, $T_s$.
\begin{align}
    \frac{dz_i}{dt} &= w^{inv}_e - (\text{Div})z_i \label{eq:inversion_height_coupled} \\
    C_s\frac{dT_s}{dt} &= S(1-\alpha) + \epsilon_c\sigma T_{b}^4 + (1-\epsilon_c)\epsilon_i \sigma T_i^4 - \sigma T_s^4 - \rho_a C_T V (\text{MSE}_s-\text{MSE}_M) - \mathrm{OHU}, \label{eq:surfaceenergy_coupled}
\end{align}
where $S$ is the diurnally-averaged insolation in June in the subtropics and $\alpha$ is the albedo of the cloud layer,
\begin{equation}
    \alpha = \alpha_c - (\alpha_c - \alpha_w)\exp{(-D\kappa \text{LWP)}},
    \label{eq:albedo}
\end{equation}
a form qualitatively consistent with \citet{gettelman2015putting}. We set a maximum cloud albedo, $\alpha_c$, and a minimum albedo representing the reflectivity of the sea surface, $\alpha_w$. The second to last term in the surface energy budget Equation \eqref{eq:surfaceenergy_coupled} is a combination of sensible and latent heat fluxes and the last term represents ocean heat uptake.

The model requires the calculation of the cloud base altitude, $z_b$, as well as the temperature of the cloud base, $T_{b}$. These may be determined by solving for $z_b$ and $T_{b}$ from the moist static energy, $\mathrm{MSE}_M = c_p T_b + L q_{tM} + g z_b$, and from the condition that   saturation is reached at the cloud base $z_b$, $q_{tM}=q^*(z_b, T_b)$, where $q^*$ is the saturation specific humidity. If $z_b$ is greater than $z_i$ (as in the decoupled, dry case), we set the cloud base equal to the cloud top to represent an infinitely thin cloud layer. At the cloud top, the specific humidity may be smaller than the total moisture, $q_{tM}$, when liquid water exists, and we similarly solve for the temperature at the cloud top, $T_{top}$, from the moist static energy, $MSE_M = c_p T_{top} + L \min\left(q_{tM}, q^*(z_i, T_{top})\right) + g z_i$, where $z_i$ is calculated using Equation \eqref{eq:inversion_height_coupled}.
 
\subsection{Diagnosing Decoupling}

To begin, we solve Equations \eqref{eq:MSE_balance_coupled}, \eqref{eq:moisture_balance_coupled}, \eqref{eq:inversion_height_coupled}, and \eqref{eq:surfaceenergy_coupled} to obtain steady-state values of MSE$_M$, $q_{tM}$, $z_i$, and $T_s$. When a stratocumulus cloud layer is coupled to the surface, kinetic energy is produced within the cloud layer by longwave cooling and entrainment drying which cools parcels such that they are negatively buoyant and sink through the mixed layer and to the surface. Decoupling occurs when negative buoyancy fluxes (downward motion of light parcels) convert turbulent kinetic energy to potential energy. The weaker turbulence weakens the convective motions and inhibits the parcel from reaching the surface \citep{bretherton1997convection}. This means that the surface is decoupled from the mixed layer. 

Following \citet{bretherton1997moisture}, we write the vertical structure of the buoyancy flux as, 
\begin{equation}
    \langle w'b'\rangle(z) = (g/s_{vo})\langle w's_v'\rangle(z),
\end{equation}
where $s_v$ is the virtual static energy (equivalent to virtual potential temperature, a buoyancy variable),
\begin{equation*}
    s_v = c_p(T + T_{ref}(0.61q_v - q_l)) + gz.
\end{equation*}
To obtain a profile of the buoyancy flux with height, we must write the virtual static energy in terms of known variables ($\text{MSE}_{M}$ and $q_{tM}$).
\begin{equation}
    s_v = \text{MSE}_M -q_vL\left(1-\frac{0.61c_p T_{ref}}{L}\right) - c_p T_{ref} q_l
\end{equation}
Decoupling occurs when there are negative buoyancy fluxes beneath the cloud base. There $q_l = 0$ and therefore $q_{tM} = q_v$. Defining $\mu = \left(1-0.61c_p T_{ref}/{L}\right)$,
\begin{align}
    s_v(z < z_b) &= \text{MSE}_M -\mu L q_{tM} \\
    \langle  w's_v'\rangle(z<  z_b) &= \langle  w'\text{MSE}_M'\rangle -\mu L \langle  w'q_{tM}'\rangle.
    \label{eq:cloud_base_BF}
\end{align}
Equation \eqref{eq:cloud_base_BF} gives the sub-cloud buoyancy flux in terms of fluxes of the known variables in the model. Rearranging Equations (\ref{eq:E_height}--\ref{eq:W_ground}), the sub-cloud buoyancy flux can be written as,
\begin{align}
    \langle w's_v'\rangle(z=z_b) &= \overbrace{(1-\frac{z_b}{z_i})C_TV(\text{MSE}_s-\text{MSE}_{M})}^{\text{Convective heat flux}} + \overbrace{\frac{z_b}{z_i}\Delta F_R/\rho}^{\text{Net longwave cooling}} \nonumber \\
    &- \overbrace{\frac{z_b}{z_i}w^{inv}_e(\text{MSE}_i - \text{MSE}_{M}) + \frac{z_b}{z_i}\mu L w^{inv}_e(q_i - q_{tM})}^{\text{Entrainment warming}} \nonumber\\
    -& \overbrace{(1-\frac{z_b}{z_i})C_TVL(q_s-q_{tM})}^{\text{Latent heat flux}} 
     - \overbrace{\mu L(F_p(z_b)-(1-\frac{z_b}{z_i})\Delta F_p) }^{\text{Sub-cloud drizzle evaporative cooling}}.
\end{align}

We invoke a strict decoupling criterion, similar to the minimal decoupling criterion in \citet{bretherton1997moisture}, such that if there are negative buoyancy fluxes beneath the cloud, the mixed layer decouples,
\begin{equation}\label{eq:DC}
    \langle w's_v'\rangle(z=z_b) \ \  
    \begin{cases}
     > 0 & \text{Coupled} \\
    \le 0 & \text{Decoupled.} 
    \end{cases}
\end{equation}
Looking at individual terms in the vertical flux profile, we can learn about the physical processes that can lead to decoupling. For example, net longwave cooling is a source of positive buoyancy flux and therefore a producer of total kinetic energy (TKE). Similarly, evaporative cooling due to cloud-top entrainment is a source of TKE. The convective heat flux at the cloud base also provides TKE by encouraging mixing. Evaporation of drizzle beneath the cloud destroys TKE by working to stabilize the stratification there. Cloud-top entrainment warming combats cooling and also destroys TKE. The competition of all of these terms determines the total sub-cloud buoyancy flux. 

We check the decoupling criterion at every time step. If it is positive, we say the mixed layer is coupled, and we accept the solution from Equations \eqref{eq:MSE_balance_coupled}, \eqref{eq:moisture_balance_coupled}, \eqref{eq:inversion_height_coupled}, and \eqref{eq:surfaceenergy_coupled}. If the decoupling criterion is negative, the mixed layer is decoupled and the well-mixed assumption is no longer valid, so we transition to the stacked mixed-layer model described below.

\subsection{Decoupled Model} \label{sec:decoupled-model}

In Figure~\ref{fig:model_schem}, we show the model configuration for a decoupled state. Instead of a single well-mixed layer, the model is split into two stacked mixed layers, a surface mixed layer (SML) and a decoupled mixed layer (DML), following \citet{turton1987study}. The DML is isolated from direct surface fluxes and receives surface moisture and energy only through mixing with the SML. Each mixed-layer has its own energy and moisture budgets, similar to Equations \eqref{eq:MSE_balance_coupled} and \eqref{eq:moisture_balance_coupled}.

The SML is coupled to the surface and experiences mixing at the top with the DML,
\begin{align}
    z_{SML}\frac{d\text{MSE}_{SML}}{dt} &= C_TV(\text{MSE}_s - \text{MSE}_{SML}) + w^{SML}_e(\text{MSE}_{DML} -\text{MSE}_{SML}) \label{eq:MSE_SML} \\
    z_{SML}\frac{dq_{SML}}{dt} &= C_TV(q_s - q_{SML}) - w^{SML}_e(q_{SML} -q_{DML}) -  z_{SML}\frac{q_s}{q_{so}}\frac{1}{\tau_\mathrm{export}}, \label{eq:moisture_SML}
\end{align}
where $w^{SML}_e$ is the entrainment rate between the SML and the DML, set as a constant in this model. We assume a constant thickness of the SML, $z_{SML}$. Note that we also include the moisture export term out of the SML as in the coupled state. 

The energy and moisture budgets for the DML are,
\begin{align}
    (z_{id}-z_{SML})\frac{d\text{MSE}_{DML}}{dt} &=  w^{inv}_e(\text{MSE}_i - \text{MSE}_{DML}) \nonumber\\
    &- w^{SML}_e(\text{MSE}_{DML} -\text{MSE}_{SML}) - \Delta F_R/\rho_a \label{eq:MSE_DML} \\
    (z_{id}-z_{SML})\frac{dq_{DML}}{dt} &= w^{SML}_e(q_{SML} -q_{DML}) - w^{inv}_e(q_{DML}-q_i) \nonumber \\
    &-  (z_{i}-z_{SML})\frac{q_{DML}}{q_{DMLo}}\frac{1}{\tau_\mathrm{export}} - \Delta F_P, \label{eq:moisture_DML}
\end{align}
where $z_{id}$ is the top of the DML, set as a constant in a decoupled case, and $z_{id}-z_{SML}$ is the thickness of the DML. The precipitation flux, $\Delta F_P$, is calculated as in Equation \eqref{eq:precip_flux}, although we note that in all the results shown in this paper there is no liquid water in the DML, so the precipitation flux is calculated to be zero.

When the mixed layer is decoupled from the surface, there is stable stratification between the SML and DML which inhibits the moisture and heat fluxes from the surface from reaching the DML. In other words, the stratification, 
\begin{equation}
    \Delta \theta_v = \theta^{SML}_v - \theta^{DML}_v > 0 \label{eq:coupling_criterion}.
\end{equation} 
At every time step, we check the sign of $\Delta \theta_v$, and if it is positive we say that the mixed layer is decoupled. When $\Delta \theta_v \le 0$, the mixed layer recouples and Equations \eqref{eq:MSE_balance_coupled}, \eqref{eq:moisture_balance_coupled} are used.

The surface energy balance is nearly identical to Equation \eqref{eq:surfaceenergy_coupled}, except now sensible and latent heat fluxes are calculated between the surface and the SML,
\begin{equation}
    C_s\frac{dT_s}{dt} = S(1-\alpha) + \epsilon_c\sigma T_b^4 + (1-\epsilon_c)\epsilon_i \sigma T_i^4 - \sigma T_s^4 - \rho_a C_T V (\text{MSE}_s-\text{MSE}_{SML})- \mathrm{OHU}. \label{eq:surfaceenergy_decoupled}
\end{equation}

\section{Results}\label{sec:results}

In this section, we present and analyze the results of the mixed-layer model for bi-stability as a function of \co{} (Section \ref{sec:results}\ref{sec:multiple-equilibria-in-co2}) and as a function of fixed-SST (Section \ref{sec:results}\ref{sec:multiple-equilibria-in-SST}). In both cases, we explore the factors that determine where the critical \co{} or SST that leads to coupling and decoupling occurs, and what is the minimum physics that allows for multiple equilibria. We do this using mechanism-denial experiments, turning off various physics and examining the results. Throughout, we emphasize new lessons learned thanks to the use of the simple model. We discuss in the conclusions the value and expected applicability of these new lessons given the simplicity of the model.

\subsection{Hysteresis in CO$_2$} \label{sec:multiple-equilibria-in-co2}

We solve Equations \eqref{eq:MSE_balance_coupled}, \eqref{eq:moisture_balance_coupled} by setting the time derivative to zero and using a root solver to find steady states of the prognostic variables as a function of \co. If the decoupling criteria (Equation \eqref{eq:DC}) is negative, we assume the mixed-layer is decoupled, and we solve Equations \eqref{eq:MSE_SML}--\eqref{eq:moisture_DML} to obtain the steady state of $q_{DML}$, $\text{MSE}_{DML}$, and $T_s$ and we check that $\Delta \theta_v$ is positive. 

\begin{figure}[!ht]
    \centering
    \includegraphics[width = \linewidth]{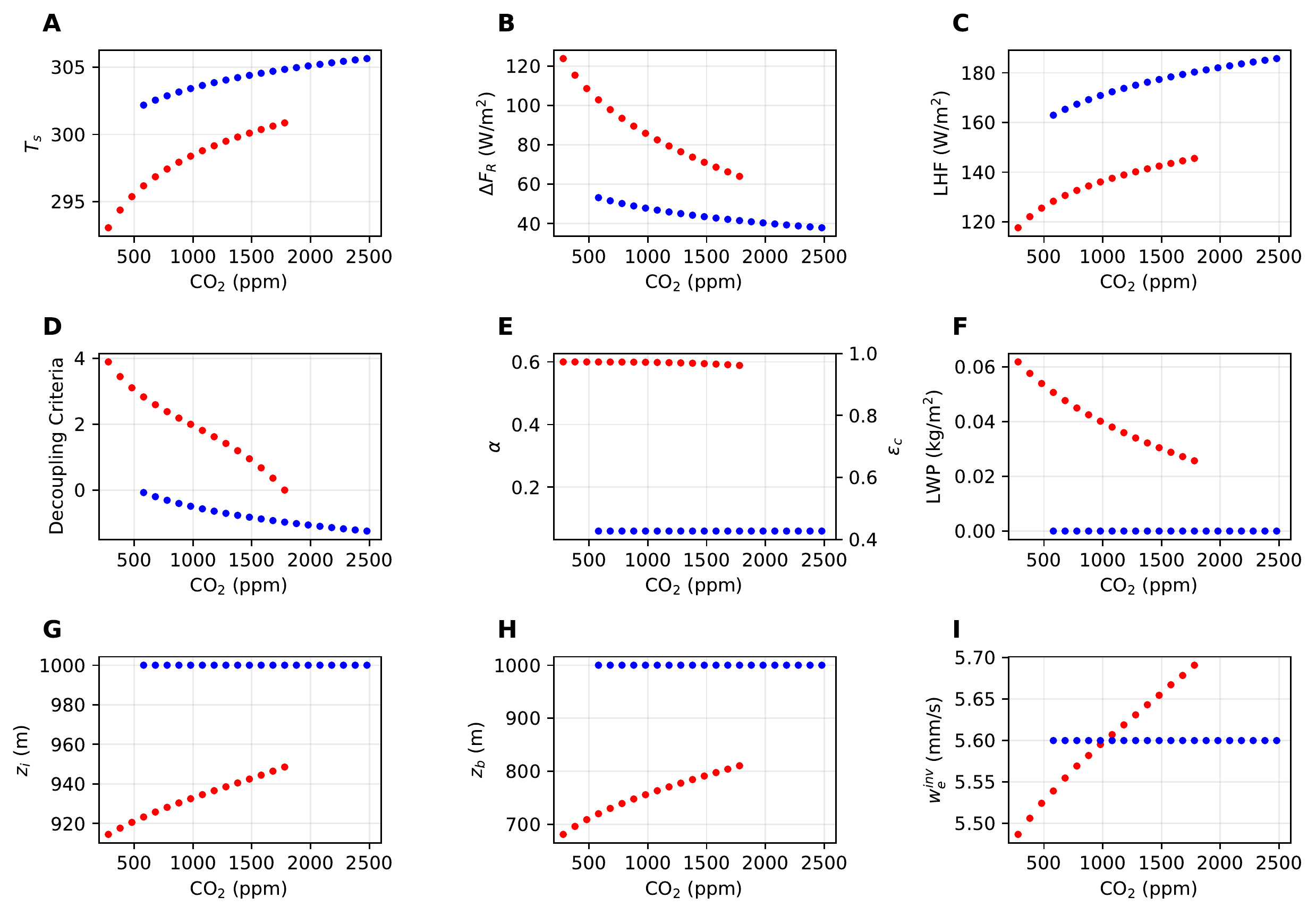}
    \caption{Hysteresis in \co. Red dots indicate coupled solutions while blue dots indicate decoupled solutions. \textbf{A.} Sea surface temperature (SST, $T_s$). \textbf{B.} Net longwave cooling across the mixed layer. \textbf{C.} Surface latent heat flux. \textbf{D.} Decoupling/coupling criteria (Equation \eqref{eq:DC} for coupled solutions and the negative of Equation \eqref{eq:coupling_criterion} for decoupled solutions).  \textbf{E.} Total albedo, including cloud and surface, and cloud emissivity. \textbf{F.} Cloud liquid water path. \textbf{G.} Height of mixed-layer top (inversion height, $z_i$). \textbf{H.} Cloud base height, where in the decoupled case when there is no cloud, the cloud base is equal to the cloud top. \textbf{I.} Cloud-top entrainment rate.}
    \label{fig:baseline_co2}
\end{figure}

Figure \ref{fig:baseline_co2} shows several variables in the \co{} hysteresis. We solve the system for \co{} values ranging from 280 ppm to 2500 ppm. Coupled solutions (plotted in red) are characterized by lower surface temperatures due to the albedo effect of the cloud layer, as in \citet{schneider2019possible}. Additionally, cloud-top longwave cooling is higher in the coupled case, driving convection and providing the cloud layer with a moisture source from the surface \citep{lilly1968models,bretherton1997moisture}. 

As \co{} increases, the emissivity of the inversion layer ($\epsilon_i$) increases, both directly from its dependence on \co{} as well as from the water vapor feedback (Equation \eqref{eq:inv_emiss}). As in \citet{schneider2019possible}, this causes a decrease in net longwave cooling across the mixed layer, $\Delta F_R$ (Figure \ref{fig:baseline_co2}B), until at around 1750 ppm, the longwave cooling is no longer strong enough to drive convection and the model predicts only decoupled solutions at higher \co.

The entrainment closure \eqref{eq:we} predicts that, all other things being equal, entrainment should weaken with longwave cooling. However, in Figure \ref{fig:baseline_co2}I, we see a slight strengthening of entrainment with \co. This is because a warmer mixed-layer leads to a weaker inversion strength, $\Delta \theta^{i-c}_v$, which allows for stronger mixing between the cloud-top and the inversion layer. 

Figure~\ref{fig:baseline_co2}D plots the decoupling criterion in red and the coupling criterion in blue. At low \co, the coupled solution predicts positive buoyancy fluxes beneath the cloud from Equation \eqref{eq:DC}, but as \co{} increases, the buoyancy fluxes beneath the cloud approach zero. Since we have enacted a strict decoupling criterion, negative buoyancy fluxes beneath the cloud lead to decoupling, so once the buoyancy flux reaches zero only decoupled solutions exist.

In the decoupled state, without a moisture source from the surface, the LWP of the DML is zero and the cloud base is equal to the cloud top (Figures \ref{fig:baseline_co2}G \& H). This causes a low albedo ($\alpha$) and cloud emissivity ($\epsilon_c$), leading to higher surface temperatures and weaker longwave cooling (Figure \ref{fig:baseline_co2}E). Once the mixed layer is decoupled, decreasing \co{} does not lead to re-coupling at 1750 ppm. Instead, \co{} must be decreased down to 500 ppm in order for the stratification between the SML and DML to become unstable and allow for convection between the surface and DML. In other words, for \co{} between 500 and 1750 ppm, there are multiple equilibria in which the layer could be either coupled or decoupled depending on its initial condition.

\begin{figure}[!ht]
    \centering
    \includegraphics[width = 0.8\linewidth]{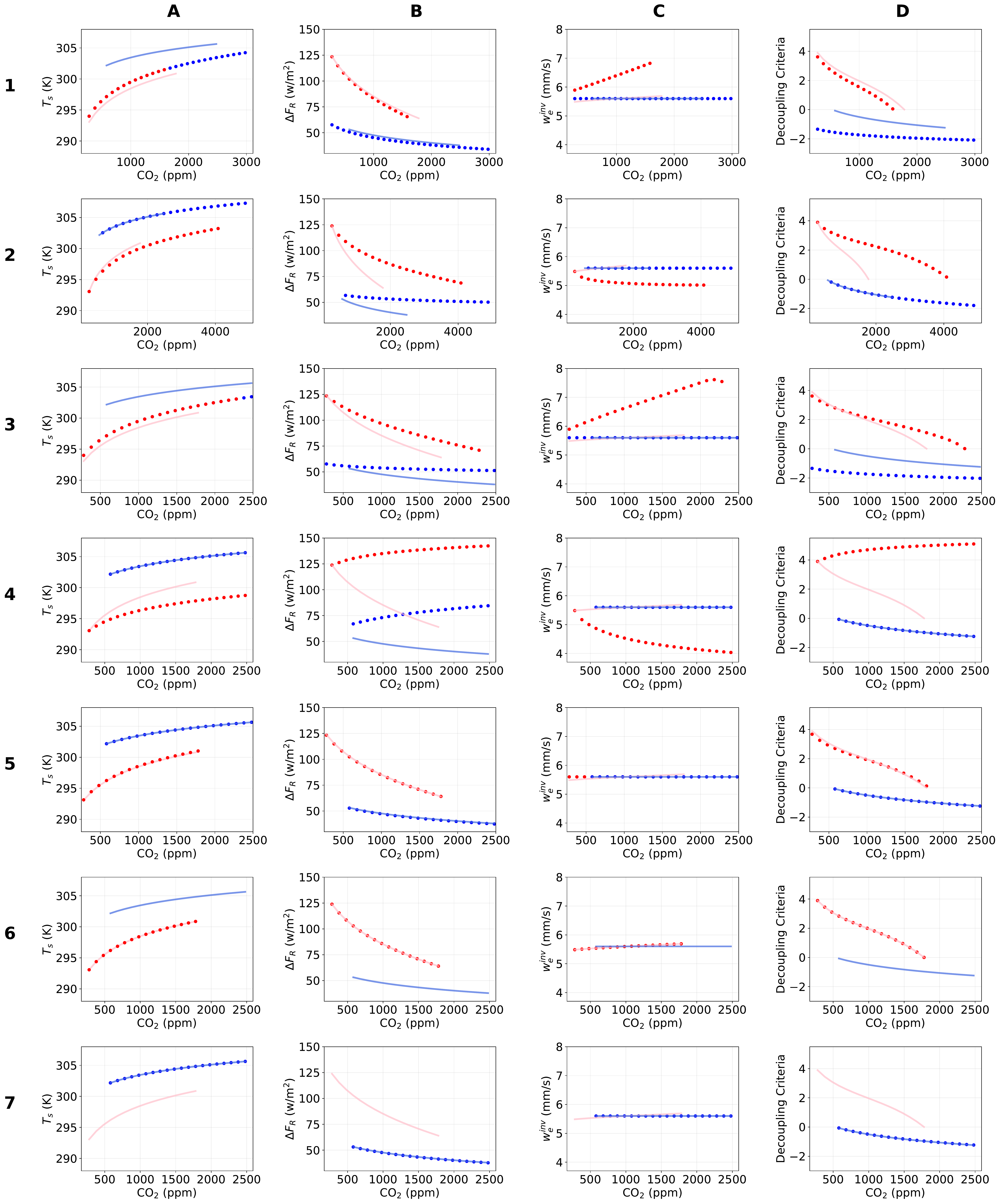}
    \caption{Summary of mechanism denial experiments for hysteresis in \co. \textbf{1.} Non-interactive SST, only a function of \co. \textbf{2.} No water vapor feedback, where inversion emissivity is only a function of \co{} (B = 0 in Equation \eqref{eq:inv_emiss}). \textbf{3.} Both 1 and 2. \textbf{4.} Constant inversion emissivity, $\epsilon_i$. \textbf{5.} Constant cloud-top entrainment, $w^{inv}_e$. \textbf{6.} Constant cloud emissivity, $\epsilon_c = 1$. \textbf{7.} Constant cloud emissivity, $\epsilon_c = 0.45$. Columns plot \textbf{A,} SST. \textbf{B,} net longwave cooling. \textbf{C,} cloud-top entrainment rate. \textbf{D,} decoupling criteria for each experiment. Transparent lines show the baseline simulation from Figure \ref{fig:baseline_co2}.}
    \label{fig:sens_co2}
\end{figure}

In general, our mixed-layer model yields similar results to those of \citet{schneider2019possible}, who used an LES. In both this model and the LES, increasing \co{} causes i) an increase in latent heat release from the surface due to enhanced evaporation, ii) a decrease in longwave cooling due to a more opaque free troposphere, iii) a decrease in LWP, iv) a decoupling of the cloud layer at high \co, and v) an abrupt increase in SST following decoupling. In their hysteresis in \co, \citet{schneider2019possible} cited the importance of amplifying cloud cover-SST feedbacks as well as the water vapor feedback for the abrupt stratocumulus cloud breakup. To test this assertion as well as explore the importance of other processes such as entrainment, we perform a suite of mechanism denial experiments in Figure \ref{fig:sens_co2}.

In the experiments shown in the first row of Figure \ref{fig:sens_co2}, we create a non-interactive SST by making $T_s$ a function of \co{} only to remove cloud cover-SST feedbacks,
\begin{equation*}
    T_s = T_{so} + 3\log_2\left(\frac{\text{CO}_2}{280}\right).
\end{equation*}
The solution is very similar to the baseline solution in Figure \ref{fig:baseline_co2}, plotted as faint lines in Figure \ref{fig:sens_co2} for easy comparison. The weakening of longwave cooling and strengthening of cloud-top entrainment are still seen with increasing \co, and decoupling occurs at high \co, with multiple equilibria. \textit{This indicates that amplifying cloud cover-SST feedbacks is not essential for decoupling in this model.}

In the experiments shown in the second row of Figure~\ref{fig:sens_co2}, we restore the interactive SST but remove the contribution of water vapor to the inversion emissivity by setting $B=0$ in Equation \eqref{eq:inv_emiss}. The most notable difference between the no water vapor feedback case and the baseline case is that it takes over twice as much \co{} for decoupling to occur. As in \citet{schneider2019possible}, at 1000 ppm, the reduction in longwave cooling is only half as much in the no water vapor feedback case than in the baseline case. However, abrupt decoupling does still occur, just at a higher \co{} than the baseline case. \textit{Therefore, in this model the water vapor feedback affects the timing of decoupling but is not necessary for decoupling to occur.}

In the experiment shown in the third row of Figure~\ref{fig:sens_co2}, we perform the above two tests simultaneously, turning off both the interactive SST and the water vapor feedback. \textit{We conclude that even with both cloud cover-SST feedbacks and the water vapor feedback disabled, abrupt stratocumulus breakup still occurs in this model.} Interestingly, \citet{schneider2019possible} cited cloud cover-SST feedbacks and the water vapor feedback as crucial components of the abrupt stratocumulus cloud break-up in their model, but in our model decoupling occurs without either of these feedbacks. This suggests that these feedbacks may not be critical for the abrupt breakup of stratocumulus clouds with increasing \co. However, we recognize that the simplicity of our mixed-layer model may mean that these processes may still be important in the parameter regime relevant to the presently observed climate. Further studies of the importance of these feedbacks should be done with more realistic models. 

In \citet{schneider2019possible}, decoupling at higher \co{} was attributed to the reduction in longwave cooling caused by the increased emissivity of the inversion layer. To test this assertion, in the fourth row of Figure~\ref{fig:sens_co2}, we set the inversion emissivity to a constant ($A = B = 0$ in Equation \eqref{eq:inv_emiss}). In panel B in the fourth row of Figure \ref{fig:sens_co2}, we see no reduction in longwave cooling which translates to no decrease in the decoupling criteria in panel D. We ran the model out to 10,000 ppm, and still, we found no critical \co{} where the model decouples and the stratocumulus clouds dissipate. \textit{This is consistent with the results from \citet{schneider2019possible} that stratocumulus breakup is caused by the increase in emissivity of the free troposphere with \co.} 

In the baseline solution, along with a weakening of longwave cooling with increasing \co, there is also a slight strengthening of the cloud-top entrainment which could also lead to decoupling \citep{bretherton1997moisture}. In the fifth row of Figure~\ref{fig:sens_co2}, we set the cloud-top entrainment to a constant and find very little difference between this test and the baseline solution, indicating that \textit{enhanced entrainment is not an important part of the dynamics for a hysteresis in \co{} in our model.} 
In all of the above-outlined sensitivity tests, the existence of multiple equilibria remains robust. In their hysteresis in SST, \citet{bellon2016stratocumulus} attributed the existence of multiple equilibria to the cloud radiative effect. That is, clouds with high LWP would have large longwave emissivity, leading to strong longwave cooling which maintains the cloud layer. Clouds with a low LWP, on the other hand, would have a small emissivity and weak longwave cooling, again maintaining the low LWP. To test this, we set the cloud emissivity in our two final mechanism denial experiments to a constant, independent of LWP. In the sixth row of Figure~\ref{fig:sens_co2}, $\epsilon_c = 1$ and in the seventh row of Figure~\ref{fig:sens_co2}, $\epsilon_c = \epsilon_{co}$. For the high (low) emissivity case, only coupled (decoupled) solutions exist because the longwave cooling is never small (high) enough to support a decoupled (coupled) solution. \textit{Therefore, in our model the dependence of cloud emissivity on liquid water path is necessary for multiple equilibria to exist, consistent with \citet{bellon2016stratocumulus}.}

In the above experiments, we see that even processes not essential for decoupling to occur can still affect the timing of decoupling. In \citet{schneider2019possible}, the critical \co{} where decoupling occurs as well as the width of the hysteresis was found to be very sensitive to model parameters such as large-scale subsidence, but the computational cost of the model precluded further sensitivity tests.

\begin{figure}[!ht]
    \centering
    \includegraphics[width = \linewidth]{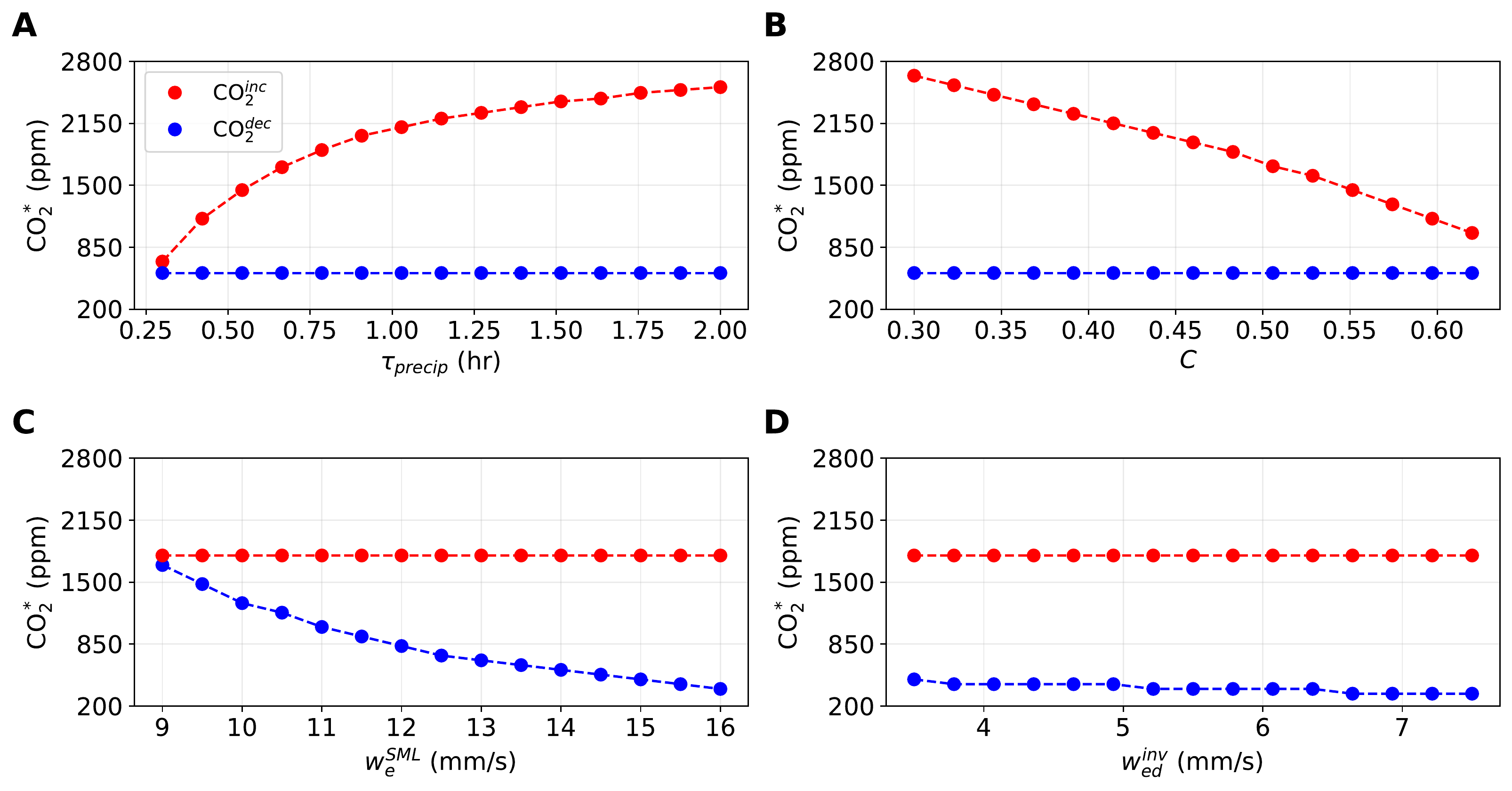}
    \caption{Critical \co{} where transition occurs when increasing \co{} (CO$^{inc}_2$) and decreasing \co{} (CO$^{dec}_2$) as a function of \textbf{A,} precipitation timescale ($\tau_{\text{precip}}$). \textbf{B,} cloud-top entrainment efficiency ($C$). \textbf{C,} entrainment between SML and DML ($w^{SML}_e$). \textbf{D,} entrainment between DML and inversion layer ($w^{inv}_{ed}$). The difference between the red and blue lines represents the width of the hysteresis.}
    \label{fig:bif_co2}
\end{figure}

In Figure~\ref{fig:bif_co2}, we explore how the width of the hysteresis (that is, the width of the SST or \co{} range for which multiple equilibria exist) varies with uncertain model parameters. Of specific interest are the precipitation timescale, $\tau_\mathrm{precip}$, and the entrainment efficiency, $C$, both of which are ill-constrained because of the complexity of cloud microphysics and aerosol-cloud interactions \citep{gerber1996microphysics, jiang2002simulations,bretherton2007cloud, wang2009modeling, ackerman2009large, uchida2010sensitivity}.

Unsurprisingly, we find that the \co{} value where decoupling occurs when increasing (CO$^{inc}_2$) is very sensitive to both of these processes. In Figure \ref{fig:bif_co2}A, decreasing $\tau_\mathrm{precip}$ (increasing the precipitation rate), leads to decoupling at a lower \co. An order of magnitude change in precipitation timescale leads to nearly a 2000 ppm increase in CO$^{inc}_2$. A decreased precipitation timescale leads to enhanced evaporative cooling beneath the cloud base which stabilizes the layer and inhibits convection \citep{stevens1998large}. Precipitation has no effect on CO$^{dec}_2$ because, in our model, LWP and therefore precipitation is zero in a decoupled case. The timescale of precipitation in reality depends on the density of cloud condensation nuclei (CCN) as well as the rate of cloud droplet merging due for example to turbulent processes \citep{Falkovich-Fouxon-Stepanov-2002:acceleration}, both of which affect the average size of the water droplet \citep{ackerman2009large}. Future work could investigate how these processes affect the width of the hysteresis in a more realistic model. 

The cloud-top entrainment efficiency, $C$, is another uncertain parameter known to play an important role in simulations of stratocumulus cloud dynamics using mixed layer models, therefore in Figure~\ref{fig:bif_co2}B, we test how $C$ affects the width of the \co-hysteresis. The model demonstrates strong sensitivity to the entrainment efficiency, showing nearly a 2000 ppm decrease of CO$^{inc}_2$ for a doubling of $C$. Once again, CO$^{dec}_2$ shows no sensitivity, since in decoupled solutions we set a constant background entrainment rate, $w^{inv}_{ed}$. For models run under prescribed SST, cloud-top entrainment from the inversion layer plays a key role in decoupling with increasing SST by stabilizing the air column through entrainment warming \citep{bretherton1997moisture}. LES have been shown to overestimate entrainment rates compared to observational best guesses, which has complicated efforts to quantify how aerosols impact entrainment \citep{bretherton2007cloud, uchida2010sensitivity}. When sedimentation feedbacks are included, high CCN concentrations lead to more efficient entrainment that decreases cloud LWP, which can cancel up to 50\% of the change in albedo due to changes in cloud droplet size \citep[Twomey effect,][]{uchida2010sensitivity}.

While our model does not include cloud microphysics, we can learn about its effects on the hysteresis and multiple equilibria indirectly from our model results, using knowledge of how CCN concentration affects precipitation and entrainment. In a parameter regime with abundant CCNs, decreasing CCN concentration should lead to an increase in precipitation  efficiency (corresponding to lower values of $\tau_\mathrm{precip}$ in Figure~\ref{fig:bif_co2}A), by increasing the average size of cloud droplets  \citep{twomey1977influence,ackerman2009large}. In our model, this corresponds to a lower CO$^{inc}_2$ as seen in Fig.~\ref{fig:bif_co2}A. Additionally, decreasing CCN concentration could lead to a decrease in entrainment efficiency (corresponding to lower values of $C$ in Figure~\ref{fig:bif_co2}B) by creating larger cloud droplets that sediment out of the entrainment zone rather than evaporating and causing evaporative cooling, and lead to less efficient entrainment \citep{bretherton2007cloud,uchida2010sensitivity}. Based on the results of our model, this could lead to an increase in CO$^{inc}_2$. The question of which process would dominate for varying CCN concentration is outside the scope of this study but would be interesting to study using more complex models.

Our treatment of the uncoupled case with two stacked mixed-layers (surface and decoupled mixed layers, denoted SML and DML, correspondingly), following \cite{turton1987study}, is highly idealized, especially with our added simplifications that entrainment between the SML and DML as well as cloud-top entrainment is constant. In Figure \ref{fig:bif_co2}C and D, we explore how sensitive the lower \co{} boundary of the multiple-equilibria region, CO$^{dec}_2$, is to these assumptions. Our model shows high sensitivity of CO$^{dec}_2$ to the entrainment rate between the SML and DML ($w^{SML}_e$) and comparatively little sensitivity to  the cloud-top entrainment ($w^{inv}_{ed}$). For decreasing \co{}, the critical value at which coupling occurs, CO$^{dec}_2$, occurs when the stratification between the SML and DML is no longer stable ($\theta_{DML} = \theta_{SML}$ at $z_{SML}$). This can be written as,
\begin{equation}
    T_{DML}\left(\frac{P_{SML}}{P_o}\right)^{R/c_p}(1 + 0.61q_{DML}) = T_{SML}\left(\frac{P_{SML}}{P_o}\right)^{R/c_p}(1 + 0.61q_{SML}),
\end{equation}
where $T_{DML}$ is the temperature in the DML at $z_{SML}$ and $T_{SML}$ is the temperature in the SML at $z_{SML}$, calculated from MSE and $q_t$ in each layer. Increasing $w^{SML}_e$ moistens the DML, such that \co{} must be decreased to very low levels in order for longwave cooling to become strong enough to decrease $T_{DML}$ and have $\theta_{DML} \le \theta_{SML}$. In an uncoupled state, the DML is very dry and therefore cloud-top entrainment does not lead to as much evaporative cooling and has a much smaller effect on the dynamics of the DML compared to mixing with the SML. This leads to a much smaller sensitivity to $w^{inv}_{ed}$ (compare Figs~\ref{fig:bif_co2}C,D). Again, CO$^{inc}_2$ does not depend on either entrainment rate since that parameterization is only included in the uncoupled state.

\subsection{Hysteresis in SST}\label{sec:multiple-equilibria-in-SST}

Previous work by \citet{bellon2016stratocumulus} used an LES to 
demonstrate how the cloud radiative effect allowed for multiple equilibria in fixed-SST. As a test for our model, we replaced the surface energy balance with,
\begin{equation}
    T_s = T_s^{fixed} ,
\end{equation}
with $T_s^{fixed}$ representing the prescribed SST with which the model is forced. \co{} is given a constant value of 280 ppm throughout this series of model experiments. In previous sections, we demonstrated that increased inversion emissivity was vital for decoupling to occur. In the model configuration used above, the inversion temperature was a function of \co{} to represent warming in the tropics due to \co{} changes. Therefore, for the fixed-SST case, we rewrite the parameterization of inversion temperature as $T_i = T_s - \Gamma_i(z_i - z_{io})$, an approximation consistent with current observed relation between the inversion and surface temperatures \citep{nicholls1984dynamics}. Once again assuming constant relative humidity, $\epsilon_i$ now increases with SST through the water vapor feedback alone.

\begin{figure}[!ht]
    \centering
    \includegraphics[width = \linewidth]{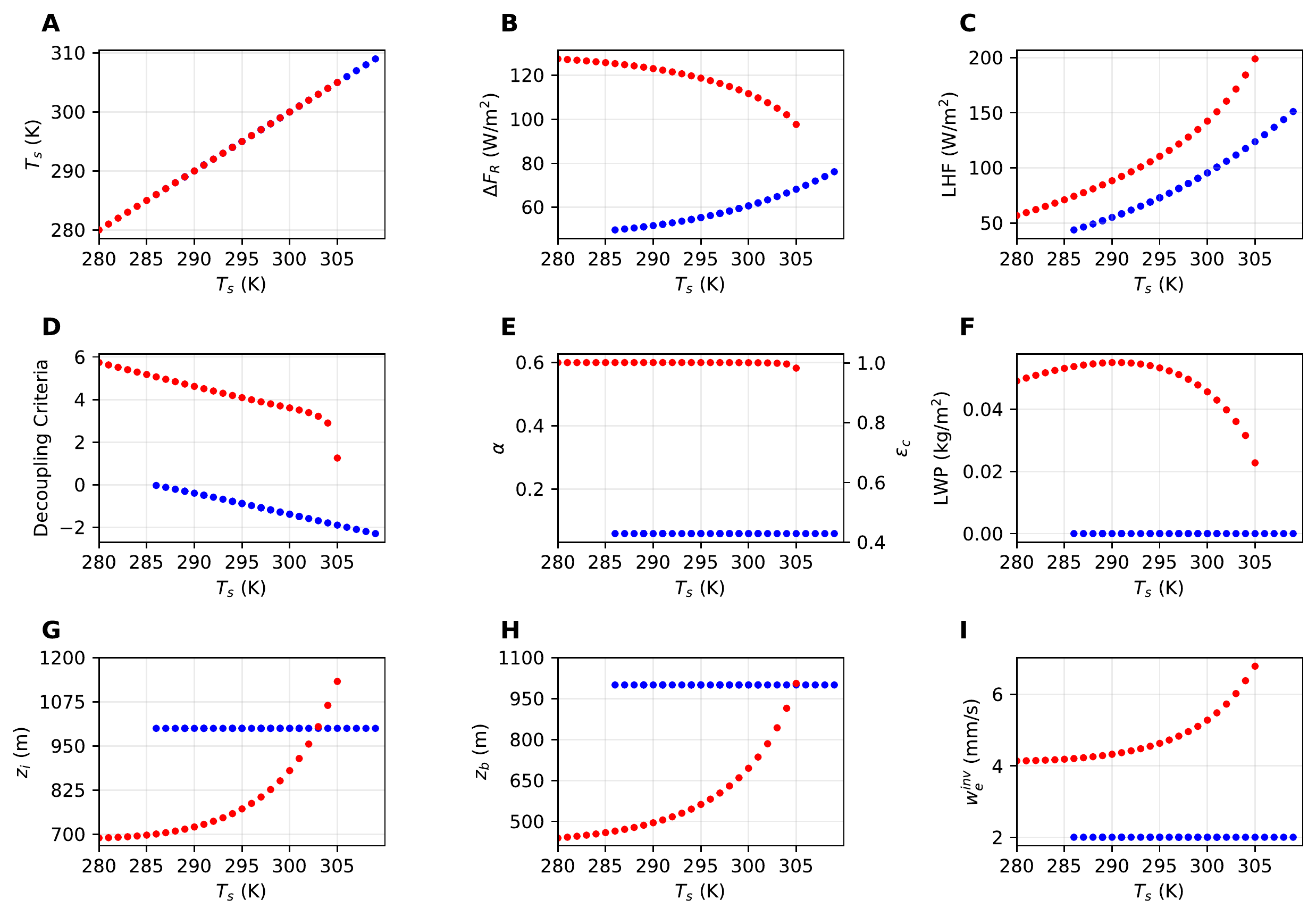}
    \caption{Hysteresis in SST. Red dots indicate coupled solutions while blue dots indicate decoupled solutions. \textbf{A.} Sea surface temperature (SST, $T_s$).  \textbf{B.} Net longwave cooling across mixed layer. \textbf{C.} Latent heat flux from surface. \textbf{D.} Decoupling/coupling criteria (Equation (\eqref{eq:DC}) for coupled solutions and the negative of Equation (\eqref{eq:coupling_criterion}) for decoupled solutions). \textbf{E.} Albedo and cloud emissivity. \textbf{F.} Cloud liquid water path. \textbf{G.} Mixed-layer depth. \textbf{H.} Cloud base. \textbf{I.} Cloud-top entrainment rate.}
    \label{fig:baseline_SST}
\end{figure}

Figure \ref{fig:baseline_SST} shows the results of our model in the fixed-SST case. Though the SST is the same for both a coupled and decoupled case, there are multiple equilibria between surface temperatures of 287 K and 305 K. In the LES study of \citet{bellon2016stratocumulus}, the SST range considered did not reveal the entire hysteresis because the low SST at which coupling occurs was not reached. Their SST range extended down to 288 K and in our model coupling occurs at 286 K, suggesting that simply expanding their SST range may have revealed the left side of the hysteresis. In addition, the critical SST where decoupling occurred was around 294 K while ours is at 305 K. In order to understand this 10 K difference, first we examine the mechanisms behind decoupling in our mixed-layer model under prescribed increasing SST.

Starting from 280 K in a coupled state, increasing SST both weakens longwave cooling (Figure~\ref{fig:baseline_SST}B) due to the water vapor feedback and strengthens cloud-top entrainment (Figure~\ref{fig:baseline_SST}I), both of which lead to a decreasing decoupling criterion (Figure~\ref{fig:baseline_SST}D). As in the LES study of \citet{bellon2016stratocumulus}, LWP has a maximum at about 292 K and then rapidly decreases due to the enhanced entrainment. The combination of enhanced entrainment and weakened longwave cooling decreases the buoyancy fluxes at the cloud base (decoupling criterion, Equation~\eqref{eq:DC}), but at 305 K decoupling happens very abruptly. This is because of the decrease in cloud emissivity (Figure~\ref{fig:baseline_SST}E) when cloud LWP dips past a threshold value set by the liquid water LW absorption coefficient, $\kappa$ appearing in Equation~\eqref{eq:ec}. This decrease in cloud emissivity results in a large decrease in net longwave cooling and results in negative buoyancy fluxes beneath the cloud base and therefore in decoupling. This effect was also noted by \citet{bellon2016stratocumulus}, who saw decoupling at 294 K. The difference from \citet{bellon2016stratocumulus} in critical SST at which decoupling occurs can be partially explained by the fact that their $\kappa$ was nearly half of ours, leading to a reduction in cloud emissivity at a higher LWP. Other differences likely exist in parameterizations of entrainment and precipitation, as well as the difference in model complexity. 

For example, \citet{bellon2016stratocumulus} prescribed the radiative effect of clouds as a function of liquid water path alone, assuming constant fluxes from the surface and free troposphere. This neglects the effect of the water vapor feedback on the emissivity of the free troposphere, which may also contribute to differences in our results. 

\begin{figure}[!ht]
    \centering
    \includegraphics[width = \linewidth]{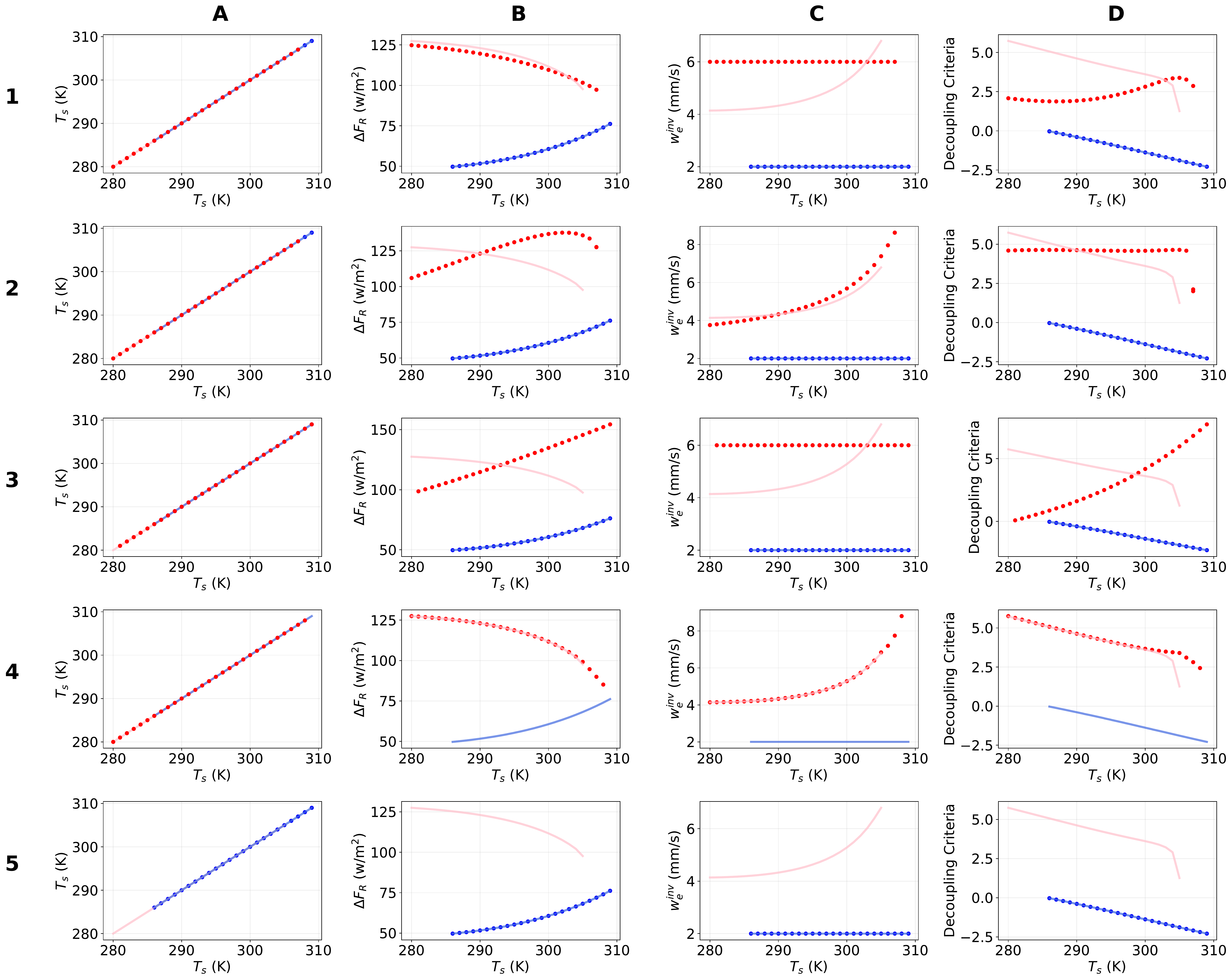}
    \caption{Summary of mechanism denial experiments for hysteresis in SST. \textbf{1.} Constant cloud-top entrainment, $w^{inv}_e$. \textbf{2.} Constant inversion emissivity, $\epsilon_i$. \textbf{3.} Both 1 and 2. \textbf{4.} Constant cloud emissivity, $\epsilon_c = 1$. \textbf{5.} Constant cloud emissivity, $\epsilon_c = 0.45$. Columns plot \textbf{A,} SST. \textbf{B,} net longwave cooling. \textbf{C,} cloud-top entrainment rate. \textbf{D,} decoupling criteria for each experiment. Transparent lines show the baseline simulation from Figure \ref{fig:baseline_SST}.}
    \label{fig:sens_all_SST}
\end{figure}

As in the analysis of the \co-hysteresis, we perform a series of mechanism denial experiments, shown in Figure \ref{fig:sens_all_SST}. As enhanced entrainment and weakening longwave cooling can both decouple the mixed layer, we begin by shutting off each of these mechanisms in turn. In the first row of Figure \ref{fig:sens_all_SST}, we set the entrainment to a constant and find that the weakening of longwave cooling due to the water vapor feedback does not prevent decoupling and only shifts the decoupling SST from 305 K to 307 K. Alternatively, in the second row of Figure \ref{fig:sens_all_SST}, setting the emissivity of the inversion layer to a constant does not inhibit decoupling either, as enhanced entrainment eventually decreases the LWP such that longwave cooling dramatically decreases. It is only when
both of these mechanisms are shut off that the mixed layer stays coupled, even up to 320 K. This indicates that in the fixed-SST hysteresis, \textit{both enhanced entrainment and weakening longwave cooling can cause decoupling on their own.} 

In both cases, decoupling happens very abruptly due to the decrease of LWP caused by both mechanisms (cloud-top entrainment and inversion layer emissivity). Even when entrainment is constant, the weakening of longwave cooling in the first row of Figure~\ref{fig:sens_all_SST} leads to a warmer mixed layer, which increases the saturation point and therefore decreases the liquid water ratio in the mixed layer. Enhanced entrainment decreases LWP both by enhanced drying but also by warming the mixed layer and allowing more water to exist in vapor form. Abrupt decoupling occurs when the LWP passes a threshold at which the emissivity of the cloud layer is no longer near-unity and longwave cooling begins to weaken. 

Finally, we test the assertion by \citet{bellon2016stratocumulus} that multiple equilibria are caused by the cloud radiative effect by setting the emissivity of the cloud to unity (experiments shown on fourth row of Figure \ref{fig:sens_all_SST}) and to the minimum value of $\epsilon_{c0} = 0.45$  (fifth row of Figure \ref{fig:sens_all_SST}), regardless of LWP. As in the \co-hysteresis, only coupled (decoupled) solutions exist when $\epsilon_c = 1$ ($\epsilon_{co}$). Once again, \textit{our model requires the dependence of cloud emissivity on LWP to show multiple equilibria, consistent with \citet{bellon2016stratocumulus}.}

\begin{figure}
    \centering
    \includegraphics[width = \linewidth]{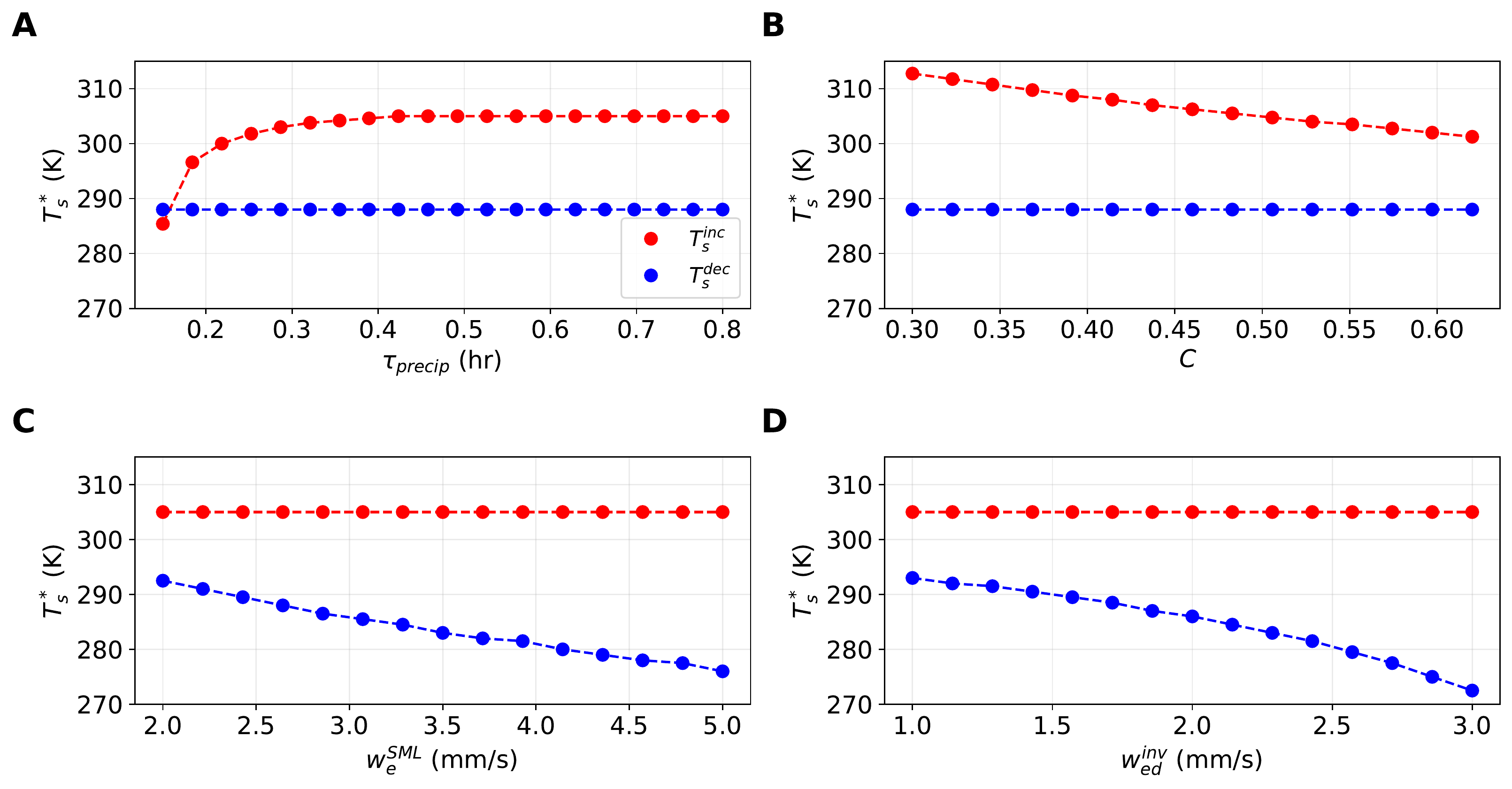}
    \caption{Critical $T_s$ where transition occurs when increasing $T_s$ ($T^{inc}_s$) and decreasing $T_s$ ($T^{dec}_s$) as a function of \textbf{A,} precipitation timescale ($\tau_{\text{precip}}$). \textbf{B,} cloud-top entrainment efficiency ($C$). \textbf{C,} entrainment between SML and DML ($w^{SML}_e$). \textbf{D,} entrainment between DML and inversion layer ($w^{inv}_{ed}$). The difference between the red and blue lines represents the width of the hysteresis.}
    \label{fig:bif_SST}
\end{figure}

We also use the computational efficiency of the model to study the sensitivity of the SST hysteresis to various model parameters in Figure \ref{fig:bif_SST}. In Figure \ref{fig:bif_SST}A, decoupling occurs at lower SST values when the precipitation timescale is very small (precipitation rate is very fast), because of the enhanced evaporative cooling beneath the cloud base. At higher $\tau_\mathrm{precip}$, the SST at which decoupling occurs, $T^{inc}_s$, asymptotes to around 305 K because at these timescales all of the precipitation evaporates before it reaches the surface, and the net precipitation flux out of the mixed layer vanishes, $\Delta F_P = 0$. In Figure~\ref{fig:bif_SST}B, we vary the entrainment efficiency, $C$, and find similarly to the \co-hysteresis case that increasing $C$ leads to decoupling at lower $T_s$ due to enhanced entrainment warming at the cloud-top. In both Figures~\ref{fig:bif_SST}A, B $T^{dec}_s$ shows no sensitivity to either parameter since there is no LWP and a constant cloud-top entrainment in the uncoupled state.

In Figure~\ref{fig:bif_SST}C, we test the sensitivity of $T^{dec}_s$ to entrainment mixing between the SML and the DML. Increasing $w^{SML}_e$ decreases $T^{dec}_s$ by moistening the DML and requiring very low surface temperatures to decrease $\theta^{DML}_v$ beyond the threshold in \eqref{eq:coupling_criterion}, similar to the \co-hystersis. In Figure \ref{fig:bif_SST}D, we test the sensitivity of the model to $w^{inv}_{ed}$. Unlike in the \co-hysteresis, there is a strong sensitivity to the cloud-top entrainment rate due to its larger size relative to $w^{SML}_e$. The hysteresis in \co{} and SST demonstrate similar sensitivities to these four model parameters. Though the prescribed SST case ignores cloud cover-SST feedbacks, it does provide an interesting case study for advected columns over warm ocean water. In both cases, precipitation and cloud-top entrainment parameterization play a large role in determining where decoupling occurs and should be studied with higher complexity models.  

\section{Conclusions}
\label{sec:conclusions}

Due to their wide coverage in the subtropics and strong albedo effect, stratocumulus clouds are an important component of Earth's energy budget. Previous work by \citet{bellon2016stratocumulus} found multiple equilibria in stratocumulus cloud cover as a function of fixed-SST, with applications to air masses advected over meridional SST gradients. Similarly, \citet{schneider2019possible} demonstrated how elevated \co{} concentrations could cause stratocumulus clouds to dissipate with multiple equilibria in cloud cover. The dissipation of the clouds corresponded to a local surface warming of up to 10 K. Both studies used a Large Eddy Simulation (LES) approach, and the model complexity precluded an in-depth study of important physical mechanisms and the sensitivity to model parameters of the critical SST or \co{} that define the range of multiple equilibria.

In this work, we developed a mixed-layer model adapted to include a surface energy balance and the ability to model decoupled solutions using ``stacked'' mixed-layers following \citet{turton1987study}. Using this model, we were able to reproduce the hysteresis as a function of  SST \citep{bellon2016stratocumulus} and \co{} \citep{schneider2019possible}. The simplicity of the model allowed us to perform mechanism denial experiments by easily switching off various physics. The computational efficiency of the model allowed us to perform a suite of sensitivity tests to find which parameters control the hysteresis behavior. 

\citet{schneider2019possible} deduced from their experiments that a dynamic SST, cloud-SST feedbacks, and the water vapor feedback were crucial for the abrupt stratocumulus cloud breakup to occur for higher \co{} values. Interestingly, \textit{we find that neither cloud-cover SST feedbacks nor the water vapor feedback are required for decoupling to occur in our model}. Instead, only an increase in inversion layer emissivity with \co{} is required. We found in extensive sensitivity experiments that \textit{the main nonlinearity needed to support the existence of multiple equilibria} either as a function of fixed SST or of \co{} \textit{is provided by the dependence of cloud emissivity on liquid water path}, as suggested by \citet{bellon2016stratocumulus} for they SST-driven hysteresis. When this nonlinearity is turned off the multiple equilibria cease to exist regardless of the many other nonlinearities that are still present. \textit{The width of the hysteresis} (that is, the range of \co{} or SST in which multiple equilibria exist) in our model also \textit{demonstrates strong sensitivity to processes which depend in reality on cloud microphysics} (precipitation and entrainment parameterizations in our model). In their study of the hysteresis in fixed-SST, \citet{bellon2016stratocumulus} attributed the decoupling and stratocumulus dissipation at higher SST to the strong enhancement of entrainment with increasing SST, but did not examine the role of the water vapor feedback. \textit{We find that decoupling with increasing SST can occur either via the enhanced entrainment or via increased inversion emissivity due to the water vapor feedback} in our model.

The model used here is highly idealized and is not meant for quantitative predictions such as the precise SST or \co{} at which stratocumulus coupling or decoupling occurs. Instead, it is designed as a tool to identify zeroth-order physical mechanisms and dependencies and to explore a wide parameter space. It is important to understand the applicability of the above-summarized lessons learned using the mechanism denial experiments based on an idealized model such as used here. A finding that a certain process is or is not essential to explain a given observation (e.g., that the water vapor feedback can cause decoupling at higher SST, or that dynamic SST is not required for the existence of multiple equilibria as a function of \co) adds to the fundamental qualitative understanding of the processes involved. A feedback that is found not to be essential for decoupling to occur, for example, may still contribute quantitatively by affecting the critical SST or \co{} at which decoupling occurs, as shown above.  The findings of the idealized model motivate further experiments with more realistic models that can then address these issues more quantitatively.

As discussed in \citet{schneider2019possible}, the stratocumulus cloud instability with \co{} presents an interesting cloud mechanism for past hothouse climates with elevated \co{} levels. This joins other cloud feedbacks such as polar stratospheric clouds \citep{Sloan-Pollard-1998:polar, Kirk-Davidoff-Schrag-Anderson-2002:feedback}, convective winter-time Arctic clouds in a warmer climate that can keep the arctic from freezing \citep{Abbot-Tziperman-2008:high}, or low clouds that can suppress polar air formation and keep winter-time continental surface temperatures above freezing \cite{Cronin-Tziperman-2015:low} that have been proposed as possible mechanisms for the Eocene equable climate period. A study of how these cloud mechanisms interact under rising \co{} levels could provide a mechanism for the equable climate problem. Such a study could benefit from simple models like the one developed and employed here.

\acknowledgment We thank Zhiming Kuang for very helpful comments and suggestions. ET thanks the Weizmann Institute for its hospitality during parts of this work. The work has been supported by the NSF Climate Dynamics program (joint NSF/NERC) grant AGS-1924538.

\newpage

\bibliographystyle{apalike}
\bibliography{References, export}
\end{document}